\documentclass[twocolumn,prl,aps,superscriptaddress]{revtex4-1}
\usepackage{braket}
\usepackage{amsmath}
\usepackage{amssymb}
\usepackage{amsfonts}
\usepackage{amssymb}
\usepackage[dvips]{graphicx}
\usepackage{subfigure}
\usepackage{dcolumn}
\usepackage{txfonts}
\usepackage{bm}
\usepackage{makeidx}
\usepackage{color}
\usepackage{mathtools}
\usepackage{threeparttable}
\usepackage[colorlinks,linkcolor=blue,anchorcolor=blue,citecolor=blue,urlcolor=blue]{hyperref}

\begin{document}

\title{Single-photon pulse induced giant response in N$>$100 qubit system}

\author{Li-Ping Yang}
\affiliation{Birck Nanotechnology Center and Purdue Quantum Science and Engineering Institute, School of Electrical
and Computer Engineering, Purdue University, West Lafayette, IN 47906,
U.S.A.}

\author{Zubin Jacob}
\affiliation{Birck Nanotechnology Center and Purdue Quantum Science and Engineering Institute, School of Electrical
and Computer Engineering, Purdue University, West Lafayette, IN 47906,
U.S.A.}

\email{zjacob@purdue.edu}
\homepage{http://www.electrodynamics.org/}

\date{\today}

\begin{abstract}
The temporal dynamics of large quantum systems perturbed weakly by a single excitation can give rise to unique phenomena at the quantum phase boundaries. Here, we develop a time-dependent model to study the temporal dynamics of a single photon interacting with a defect within a large system of interacting spin qubits (N$>$100). Our model predicts a new quantum resource, giant susceptibility, when the system of qubits is engineered to simulate a first-order quantum phase transition (QPT). We show that the absorption of a single photon pulse by an engineered defect in the large qubit system can nucleate a single shot quantum measurement through spin noise read-out. This concept of a single-shot detection event (``click") is different from parameter estimation which requires repeated measurements. The crucial step of amplifying the weak quantum signal occurs by coupling the defect to a system of interacting qubits biased close to a QPT point. The macroscopic change in long-range order during the QPT generates amplified magnetic noise, which can be read out by a classical device.  Our work paves the way for studying the temporal dynamics of large quantum systems interacting with a single-photon pulse.
\end{abstract}

\maketitle
\textbf{Introduction}\\
Recent developments in controlling large quantum systems  in cold atoms systems\cite{bernien2017probing}, ion traps~\cite{zhang2017observation}, and superconducting qubit systems\cite{Harris2018Phase} have opened the new era of quantum simulation. In particular, the study of continuous quantum phase transitions and many body localization promises to be one of the major applications for quantum computers~\cite{arute2019quantum}. Simultaneously, control over large quantum systems allows sensing and parameter estimation with unprecedented sensitivity. In particular, continuous quantum phase transitions combined with repeated measurements can be exploited as a resource for metrology~\cite{quan2006decay,zanardi2008quantum}. This opens the question whether a giant response can occur in a large quantum system even when weakly perturbed by a single photon~\cite{lv2018single}. Such a system with a giant response can lead to single shot read-out without the need for repeated measurements. 


We develop a time-dependent computational model to study the  response of a large quantum system ($N>100$ qubits) on excitation by a single photon pulse. We discover a giant response arising from a single photon interacting with a defect state coupled to a large system of collective qubits ( N$>$100), which can function as a quantum amplifier~\cite{yang2019QCD}. We believe this striking giant response will motivate experiments of the time dynamics of large quantum systems excited by a single-photon pulse. While we capture the essential physics through a minimalistic model, it points to single photon nucleated space-time theory of quantum phase transitions where even excited states along with ground states play an important role. This can lead to an exciting frontier at the interface of condensed matter physics and quantum optics. Our proposal can be implemented in a broad range of qubit systems and can lead to devices such as single photon detectors in spectral ranges inaccessible by current technologies. 

\begin{table*}
\centering
\begin{tabular}{|p{150pt}<{\centering}||p{150pt}<{\centering}||p{150pt}<{\centering}|}
\hline 
Quantum  entanglement/squeezing & Quantum criticality &  \textbf{Giant quantum susceptibility}\\
\hline {Based on GHZ~\cite{Bollinger1996optimal} or squeezed spin states~\cite{wineland1994squeezed}} & {Based on second-order (continuous) quantum phase transition~\cite{quan2006decay,zanardi2008quantum}} & {Based on first-order (discontinuous) quantum phase transition}\\
\hline {Heisenberg uncertainty principle} & {Orthogonality of the ground states of two neighbouring quantum phases} & {Discontinuous change in the long-range  spin order }\\
\hline
{High phase sensitivity $\partial P/\partial \phi\propto N$ (slope of the population $P$ with respect to the phase $\phi$)~\cite{giovannetti2004quantum}} & {High fidelity susceptibility $\partial \mathcal{L}/\partial\lambda\propto N$ (slope of the Loschmidt echo $\mathcal{L}=|\langle G(\lambda)|G(\lambda+\delta\lambda)\rangle|$)} & {Giant quantum susceptibility $\partial |M|/\partial \lambda\propto N$ (slope of the spontaneous magnetization $|M|$)}\\
\hline 
{Phase interference based and repetitive measurements required~\cite{giovannetti2004quantum}} & Repetitive measurements of the decoherence of an ancillary qubit~\cite{quan2006decay} & {Single-shot measurement; Non-adiabatic transitions dominate dynamics}\\
\hline
\end{tabular}
\vspace{0.2cm}
\caption{\label{tab1} We contrast three fundamental classes of quantum resources. Giant quantum susceptibility proposed in this paper against previously well-known quantum entanglement/squeezing~\cite{wineland1994squeezed,Bollinger1996optimal,giovannetti2004quantum} and quantum criticality~\cite{quan2006decay,zanardi2008quantum}. }
\end{table*}

\begin{figure*}[hbt!]
\vspace{0.4cm}
\centering
\includegraphics[width=15cm]{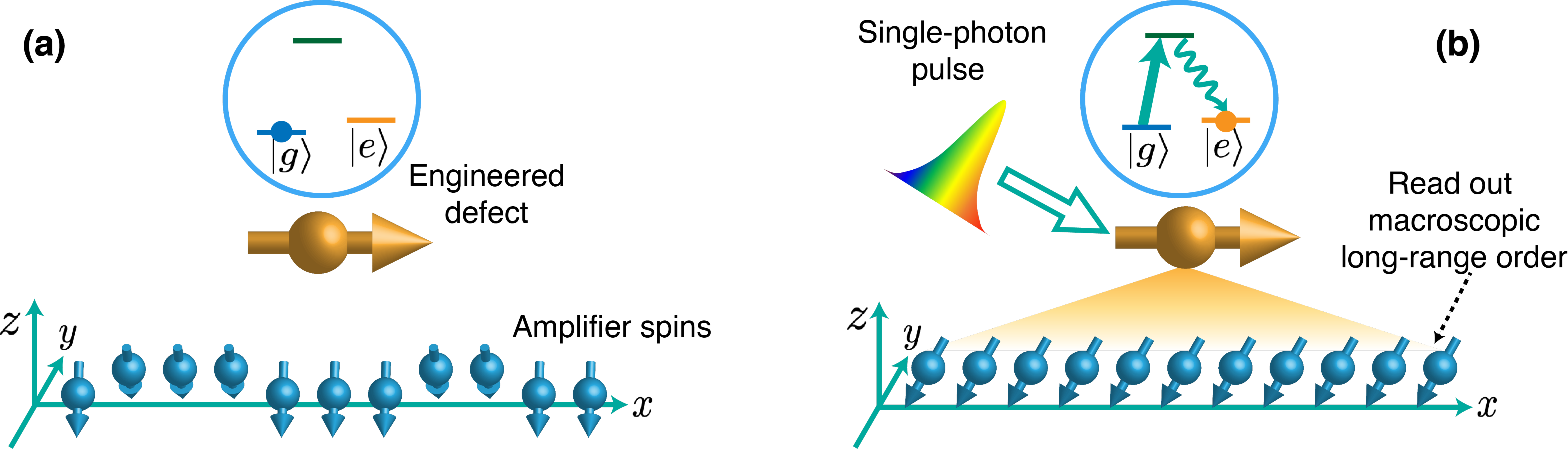}\caption{\label{fig:1} \textbf{Computational model of single photon pulse interacting with $N>100$ qubit system} The interacting spin qubits at the bottom, which can function as a quantum amplifier, are critically biased close to the first-order QPT point. The three states in the absorber on the top form a $\Lambda$-structure. After absorption of a single-photon pulse, the absorber is excited from the ground state $|g\rangle$ and finally relaxes to the meta-stable state $|e\rangle$. After the $|g\rangle\rightarrow |e\rangle$ transition, the absorber exerts an effective magnetic field on the amplifier qubits. This magnetic field triggers a QPT in the qubits underneath. Initially, the spin qubits are polarized in the $yz$-plane (see panel \textbf{a}). After the phase transition, the spins rotate to the $xz$-plane (see panel \textbf{b}).}
\end{figure*}

We show that giant susceptibility is a new quantum resource and we overcome two outstanding challenges for the field. Firstly, previous proposals of metrology exploiting second-order QPTs (see Table~\ref{tab1}) only give rise to large fidelity susceptibility, not a directly observable quantity. The quantum susceptibility is always low for previously studied second-order QPTs making it detrimental for practical experimental realization. Existing schemes propose to use repetitive measurements on a weak output signal to perform high precision parameter estimation i.e. quantum metrology - fundamentally different from our claim of single photon pulse driven giant response. For example, in the transverse Ising model, the magnetic susceptibility $\chi$ diverges at an extremely low speed with the spin number $N$ ($\chi\sim log(N)$)~\cite{um2007quantum}. Thus, no giant response can be obtained for intermediate scale quantum systems. On the contrary, the susceptibility diverges with $N$ at the first-order QPT point in our proposed model. Secondly, the time dynamics of a weak signal (e.g.: a single photon pulse) interacting with any large system near a quantum phase transition has never been explicitly demonstrated. We overcome this challenge by exploiting recent developments in quantum pulse scattering theory~\cite{yang2018concept,yang2019single,Baragiola2012photon,Wang2011efficient}. 

Exploiting giant quantum susceptibility is an approach fundamentally different from  quantum interferometers used for parameter estimation or quantum sensing/metrology~\cite{Degen2017quantum,demkowicz2012elusive,giovannetti2004quantum} (see Table~\ref{tab1} for a summary of these differences). The enhanced sensitivity in quantum interferometers benefits from the accumulated phase from a large number of synchronized non-interacting particles in repeated measurements~\cite{Zhang2018noon,Davis2016approaching,Thorne2018nobel}.  In contrast, the giant sensitivity in our scheme originates from the singular behavior of strongly correlated systems at the phase transition point~\cite{yang2019QCD,yang2019engineering}. This giant response can give rise to classical single-shot measurements. 

The dynamic process of the proposed interacting qubit system is conceptually similar to the counting events in single photon avalanche diodes (SPADs) and SNSPDs. This is clarified on contrasting our approach with the well-established and important field of quantum linear amplifiers~\cite{Caves2012quantum,bergeal2010analog}. The gain of linear quantum amplifiers arises from the coherent pumping in the ancillary modes. Simultaneously, phase information is encoded in the quadratures of the signal modes which is preserved during the amplification. However, the quantum gain of our system results from the macroscopic change in the order parameter during the QPT. The phase information in the input signal (e.g., the pulse shape) is lost during the amplification and only the pulse number information ($0$ or $1$) is read out by the amplifier. 


\begin{figure*}
\centering
\includegraphics[width=12cm]{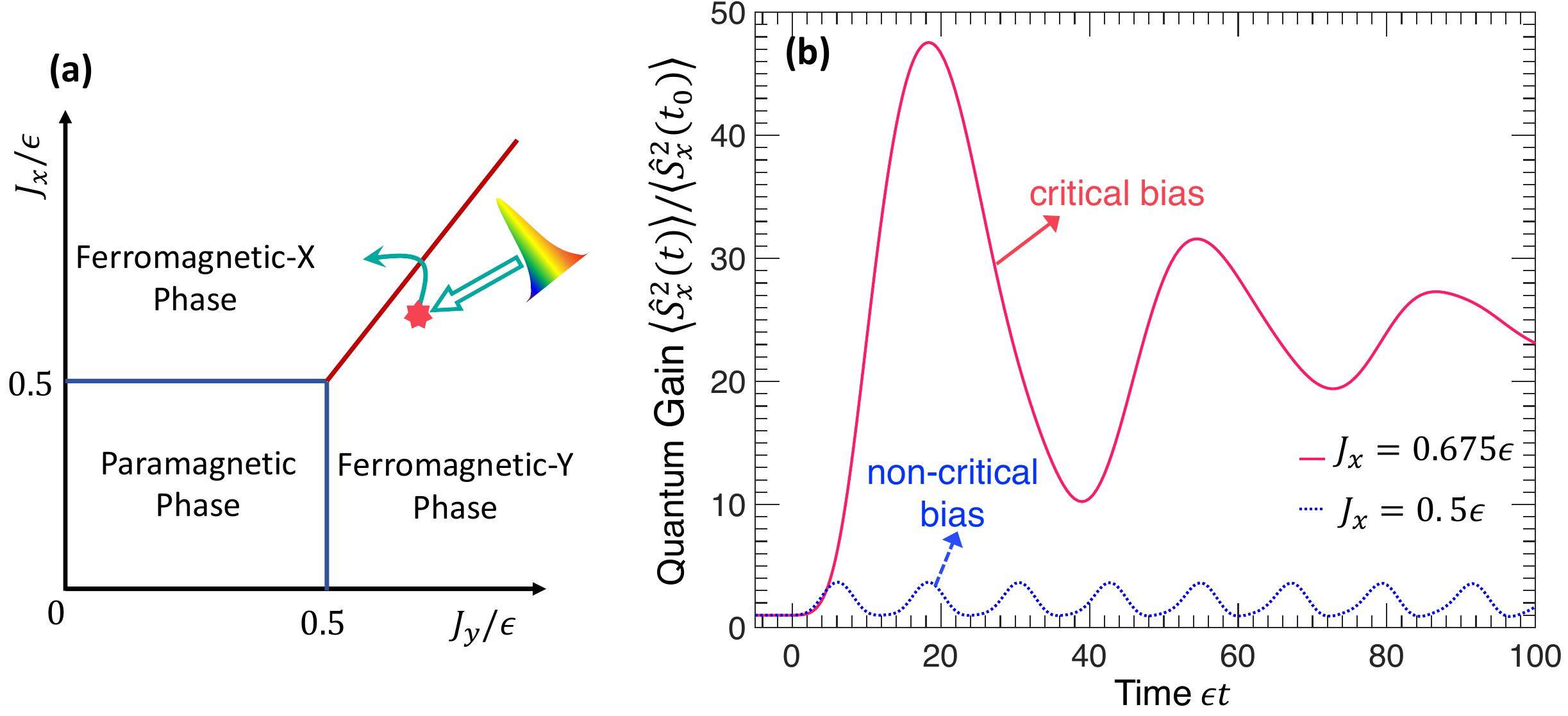}\caption{\label{fig:2} \textbf{Single-photon pulse induced first-order QPT. a} The phase diagram and the initial bias of the amplifier. \textbf{b} After absorption of a single-photon pulse, the absorber is flipped to the state $|e\rangle$, on which the absorber exerts a weak magnetic field $B_x\times P_e(t)$ ( $P_e(t)$ the population of the state $|e\rangle$) on the amplifier with $N=400$ qubits. Only if the the amplifier is optically biased around the critical point $J_{x,c}\equiv J_y$, the time varying field can trigger a first-order QPT to obtain a large quantum gain. Here, the qubit-qubit
coupling in the $y$-direction is fixed at $J_y=0.7$ and the absorber-amplifier coupling $B_x=0.01\epsilon$. }
\end{figure*}

\textbf{Results}\\
\textbf{Single-photon detection}---We now discuss the working mechanism and implementation of our model. The first step is the transduction (absorption) of the incident single photon in an engineered defect. This process is similar to the generation of the first electron-hole pair in single-photon avalanche diode or the first photo-emission event in the photo-multiplier tube. The highly efficient transduction is realized via a $\Lambda$-structure transition as shown in Fig~\ref{fig:1}. In contrast to a two-level absorber, this $\Lambda$-transition defect has three main benefits: (1) higher absorption probability~\cite{young2018fundamental,Wang2011efficient}; (2) longer lifetime of the destination state $|e\rangle$~\cite{yang2018concept} conducive for effecient read-out; (3) connection of the optical transition in the absorber and the RF-frequency dynamics in the interacting-qubit system (the amplifier).  One promising example of such kind of absorber is a nitrogen-vacancy (NV) center.  The states $|g\rangle$ and $|e\rangle$ correspond to the two ground spin states $|0\rangle$ and $|+1\rangle$ of the NV. The $T_1$ time (lifetime of the state $|e\rangle=|+1\rangle$) of NV centers is few milliseconds at room temperature and even much longer at lower temperatures~\cite{Jarmola2012temperature}. The $\Lambda$ transition can be realized with the spin non-conserving transition~\cite{Chu2015all} as shown in the supplemental material~\cite{supplementary}. After the transduction, the information of the single-photon pulse is written in the $|e\rangle$ state of the absorber.

Amplification is essential to trigger a giant response, since the signal stored in the defect (absorber) after the transduction is usually an extremely weak quantum signal. In our quantum system, the amplifier consists of a large number of interacting ancilla qubits. An important principle is effective engineering of the absorber-amplifier interaction to guarantee that the absorbed energy is transferred to the readout channel to trigger the QPT. In our model, the coupling between the absorber and the amplifier is engineered in $x$-direction
\begin{equation}
\hat{H}_{\rm int}=B_x \left|e\right\rangle \left\langle e\right|\sum_{j}\hat{\sigma}_{j}^{x}.\label{eq:dispersive_coupling}
\end{equation}
This dispersive coupling with strength $B_x$ acts an effective magnetic field for the amplifier qubits. As shown in the following, the defect functions as a control of the QPT in the amplifier. More importantly, the dispersive coupling avoids additional decoherence of the amplifier induced by the single-photon pulse. The NV center couples to its surrounding spin qubits dispersively as in equation (\ref{eq:dispersive_coupling}) when the strength $B_x$ is much smaller the ground-state zero-field splitting $\Delta_{\rm gs}\approx 2.87$~GHz~\cite{zhao2011atomic}.

In our proposal, the amplification is realized by exploiting the giant sensitivity of the first-order QPT. With the mean-field theory, we predicted a universal first-order QPT in interacting qubit systems~\cite{yang2019engineering}
\begin{equation}
\hat{H}_{\rm Am}=\frac{1}{2}\epsilon\sum_{j=1}^{N}\hat{\sigma}_j^{z}-\frac{1}{n}\sum_{\langle i<j \rangle}(J_{x}\hat{\sigma}_i^x\hat{\sigma}_j^x + J_{y}\hat{\sigma}_i^y\hat{\sigma}_j^y), 
\end{equation}
where $\epsilon$ is the energy splitting of the qubits along $z$ direction, $J_x$ and $J_y$ are the strengths of the ferromagnetic qubit-qubit couplings in $x$- and $y$-direction respectively, and $\hat{\sigma}_j^{\alpha}\ (\alpha = x,y,z)$ are the Pauli matrices of the $j$th qubit. The summation $\langle i<j \rangle$ runs over $n$ coupled neighbours. For the 1-dimensional Ising chain with~$n=1$, the short-range coupling only exists between the nearest neighbours~\cite{sachdev2007quantum}. For the Lipkin-Meshkov-Glick (LMG) model with $n=N-1$ ($N$ the total qubit number)~\cite{lipkin1965validity}, all the qubits are coupled with each other. 

The amplifier qubits has two ferromagnetic (FM) phases: FM-X and FM-Y with long-range spin order in $x$- and $y$-direction. The competition between these two FM phases results in the first-order QPT, which exhibits giant sensitivity for weak signal detection~\cite{yang2019engineering}. In Fig.~\ref{fig:2} (a), we present the schematic of the first-order QPT boundary (the red line) in the phase diagram. The quantum phases and the corresponding QPTs can be characterized by two magnetic order parameters
\begin{equation}
\zeta_{x}\equiv \langle \hat{S}_x^2\rangle_0/N^2\ {\rm and}\ \zeta_{y}\equiv \langle \hat{S}_y^2\rangle_0/N^2,      
\end{equation} 
which describe the magnetic fluctuations in the $xy$-plane. Here, $\hat{S}_{\alpha} = \sum_j\hat{\sigma}_j^{\alpha}/2$ are the collective qubit operators and $\langle\cdots\rangle_0$ means average on the ground-state of the amplifier. The second-order QPTs in interacting qubit systems have been extensively demonstrated in recently experiments~\cite{zhang2017observation,bernien2017probing,Harris2018Phase}. Specifically, the second-order QPT in the LMG model (with long-range qubit-qubit coupling only in $x$-axis) has also been demonstrated in a recent experiment with $16$ Dysprosium atoms~\cite{makhalov2019probing}. We suggest that by adding an additional laser to induce the long-range coupling in $y$-direction, the first-order QPT due to the competition between the two FM phases can also be observed. This can provide a promising platform to build a single-photon detector utilizing first-order QPT in the LMG model. In the following, we numerically demonstrate the single-photon pulse induced giant response near the first-order QPT of the LMG model, which occurs at $J_x=J_y>\epsilon/2$~\cite{yang2019engineering}.

The amplification and single-shot readout of the quantum information stored in the state $|e\rangle$ is realized by exploiting the first-order QPT in the amplifier. Initially, the qubit-qubit coupling $J_x$ is pre-biased slightly below the phase transition point $J_{x,c}\equiv J_y$ [see the red star in Fig.~\ref{fig:2} (a)] and the amplifier is initialized in its ground state of the FM-Y phase. After absorption of a single-photon pulse, the absorber is flipped to the state $|e\rangle$ with probability $P_{e}(t)$ [see supplemental material~\cite{supplementary}]. Thus, the additional effective magnetic field experienced by the amplifier qubits is $B_x\times P_{e}(t)$.  The initial critical bias guarantees that the small magnetic field perturbation $B_x\times P_{e}(t)$ from the absorber can trigger a QPT and leads to efficient amplification. 

\begin{figure*}
\includegraphics[width=16cm]{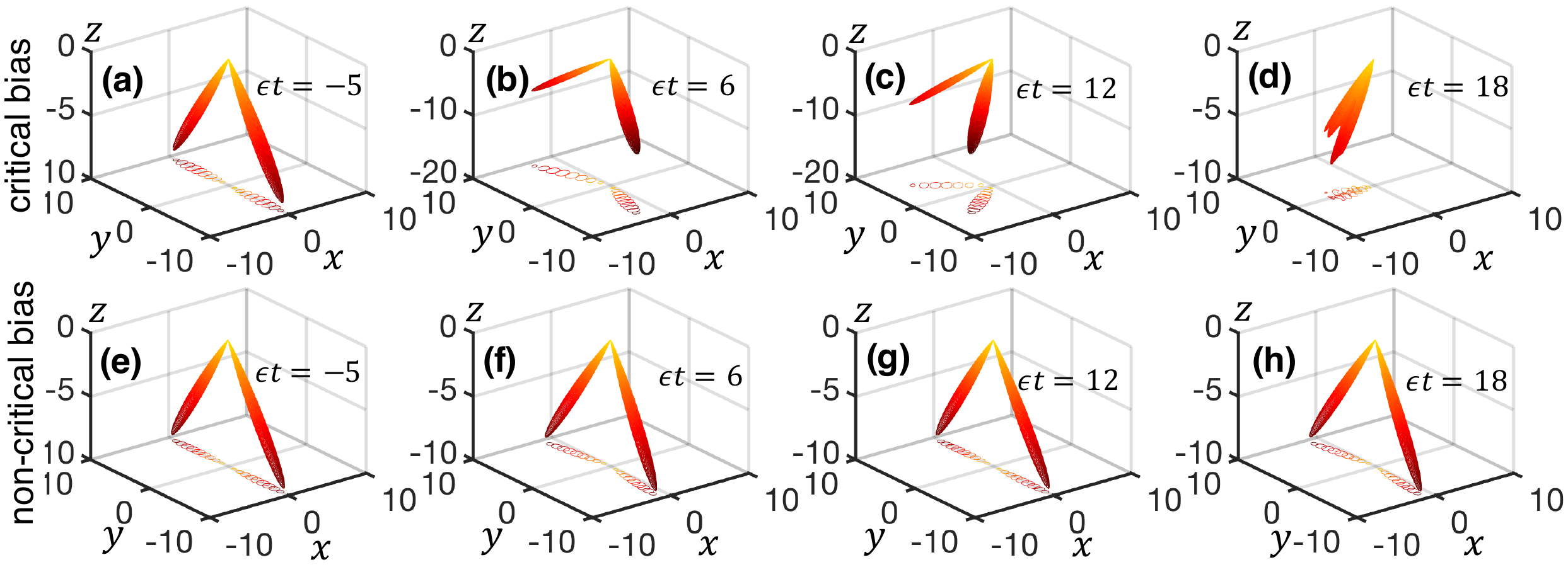}
\centering
\caption{\label{fig:3} Dynamics of the amplifier. The spin $Q$-function characterizes the polarization distribution of the amplifiers qubits.  The first rows (\textbf{a-d}) shows the dynamic change in the $Q$-function with bias $J_x =0.675\epsilon$ very close to the phase transition point $J_{x,c}=J_y=0.7\epsilon$. The second row (\textbf{e-h}) is for the case with $J_x=0.5\epsilon$ far from $J_{x,c}$. The curves underneath are the contour projections of the corresponding $Q$-functions in $xy$-plane.}
\end{figure*}

There are two ways to read out the amplified signal in practice. One is to directly measure the spontaneous magnetization $\sqrt{\zeta_x}$ of the amplifier in $x$-direction, which increases from an extremely small value to a finite value after the collective rotation of the qubits. Another option is to couple the amplifier qubit with a cavity as proposed in our previous works~\cite{yang2019QCD,yang2019engineering}. The energy prestored in the qubits is transferred to the cavity mode generating macroscopic excitations after the QPT. The photons leak out from cavity can be directly measured with classical photodetectors. 

To characterize the dynamic giant response, we define a time-dependent quantum gain of the amplifier as
\begin{equation}
G(t)=\langle\hat{S}_x^{2}(t)\rangle/\langle\hat{S}_x^{2}(t_0)\rangle.
\end{equation} 
We contrast the time-dependent quantum gain for the cases of critical bias (the red-solid curve) and non-critical bias (the blue dotted curve) in Fig.~\ref{fig:2} (b). It is clearly seen that the giant response (corresponding to an efficient amplification) can only be obtained if the system is optimally biased close to the phase transition point~\cite{yang2019QCD}. We also note that for the critical bias case, the amplifier qubits finally evolve to an excited state in the FM-X phase with macroscopic qubits polarized in $xz$-plane as shown in following. 

To reveal the intrinsic change within the amplifier, we contrast the time-dependent spin $Q$-function of the amplifier for different biases in Fig.~\ref{fig:3}.  The first row (a-d) and the second row (e-h) correspond to critical and non-critical bias cases, respectively. In both cases, the amplifier starts from the FM-Y phase with spin qubits polarized in the $yz$-plane. The two arms of the $Q$-function in the $yz$-plane at time $t_0=-5/\epsilon$ (the time before the absorption of the pulse) correspond to the two degenerate ground states of the FM-Y phase~\cite{yang2019engineering}. For the first row, the incident single-photon pulse triggers a phase transition to the FM-X phase. The qubits rotate $90{\rm ^o}$ to the $xz$-plane at time $t_0=18/\epsilon$ in Fig.~\ref{fig:3} (d). This reveals the dynamic change in the long-range spin order within the amplifier and clearly shows the signature of the detection event. In contrast, no macroscopic spin order change occurs when the amplifier is biased far from the phase transition point. The polarization of the spin qubits marginally varies with time in Fig.~\ref{fig:3} (e-h).  

Our simulation of the amplifier dynamics has ignored the decoherence of the interacting qubits that may degrade the giant response in practical processes. However, the amplification has completed within the time $\epsilon T_{\rm Am}\approx 15$~\cite{supplementary}, which is usually much shorter than the decoherence time of the qubits. If the amplifier is composed of electron spins with typical energy splitting $\epsilon\sim 1$~GHz and coherence time $T_2^{*}\sim 1{\rm \mu s}$~\cite{Ryan2010robust}, we have $\epsilon T_2^{*}\approx 1000\gg \epsilon T_{\rm Am}$. For nuclear spins with typical energy splitting $1$~MHz and coherence time $T_2^*\sim 1$~ms at room temperature and longer than $10$s at low temperature~\cite{yang2016high}, the decoherence time is still much longer than the amplification time. With dynamical decoupling techniques~\cite{zhao2012decoherence,yang2016quantum}, the coherence time of the spins can be further prolonged $2-3$ orders of magnitude~\cite{balasubramanian2009ultralong,Ladd2005coherence,maurer2012room}, which is far more than the required time for amplification. The dipole-dipole interaction between the NV center and nuclear spins at the typical distance $1$~nm is around $20$~kHz. This effective magnetic field ($B_x/\epsilon\approx 0.02$) is large enough to trigger the QPT.

\begin{figure}
\centering
\includegraphics[width=8.5cm]{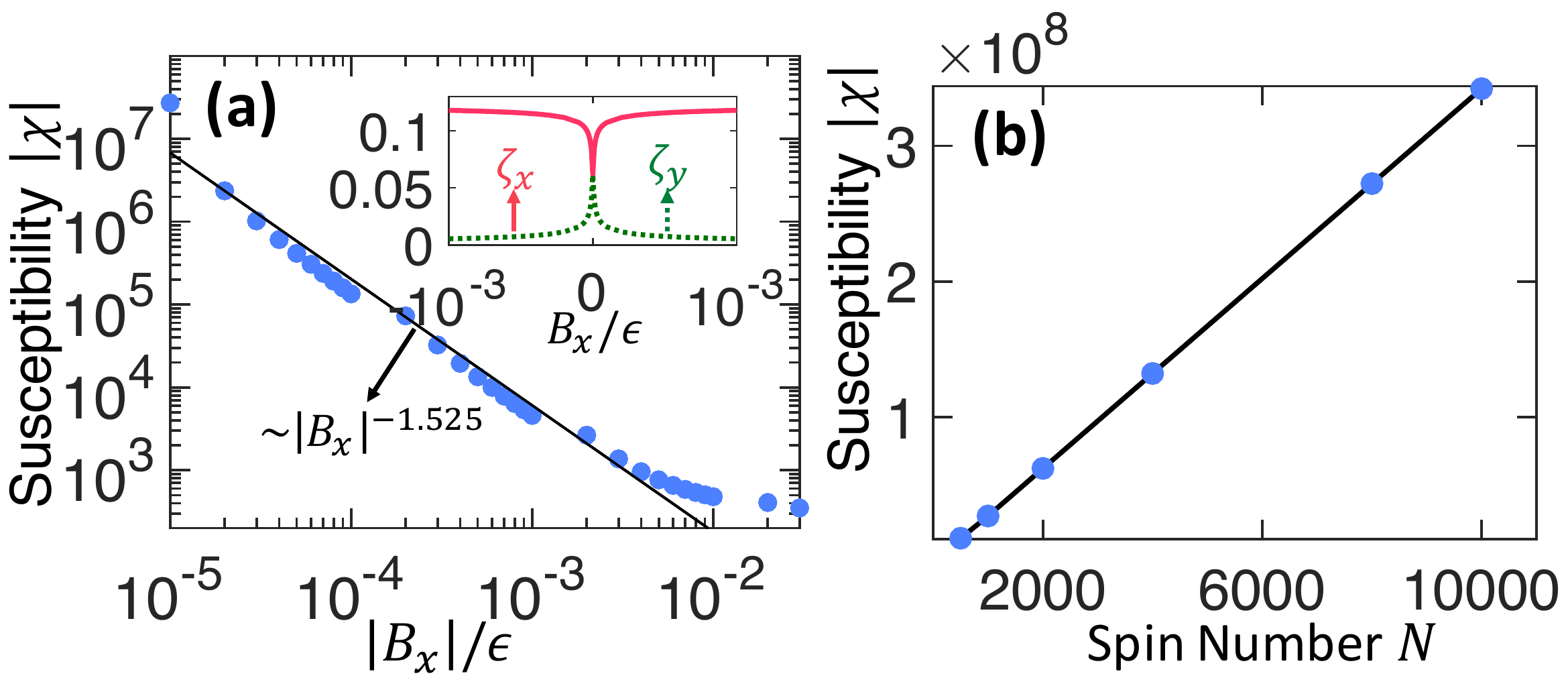}\caption{\label{fig:4} Singular behavior in the susceptibility. (a) The susceptibility $\chi$ diverges at the phase transition point $J_x=J_y=0.7\epsilon$. The subgraph shows the abrupt changes in the order parameters in the first-order QPT transition with qubit number $N=1000$. (b) The susceptibility $\chi$ near the phase transition point increases linearly with the qubit number $N$. Here, the perturbation magnetic field is set as $B_x=10^{-5}\epsilon$.}
\end{figure}

\textbf{Discussion}\\
\textbf{The singular scaling of the interacting qubit system}---The giant response of the interacting qubits fundamentally originates from the singular behaviors of the system at the phase transition point. We now show the singular scalings of the system. We also notice that in most cases, it is difficult for weak input signals to change the coupling strength within the amplifier~\cite{yang2019QCD}. Here, we show that a weak magnetic field perturbation can also break the balance of the two FM phase at the phase boundary $J_x=J_y$ to trigger the first-order QPT. This also lays the foundation of the amplification mechanism as shown in the previous section. As shown in the subgraph of Fig.~\ref{fig:4} (a), the order parameter $\zeta_x$ increases swiftly with the perturbation magnetic field in $x$-direction and the other order parameter $\zeta_y$ drops. The sensitivity to the magnetic field is characterized by the susceptibility of the spontaneous magnetization
\begin{equation}
\chi=\left.\frac{d\sqrt{\zeta_x}}{dB_x}\right|_{B_x\rightarrow0}\propto |B_x|^{-\gamma},   
\end{equation}
which is symmetric on the two sides of the transition with singular exponent $\gamma\approx 1.525$. The same susceptibility for the spontaneous magnetization $\sqrt{\zeta_y}$ with respect to a magnetic field in $y$-axis can also be obtained  (data not shown). The susceptibility diverges linearly with the qubit number $\chi\sim N$ as shown in Fig.~\ref{fig:4} (b).

We emphasize that in first-order QPTs, a singularity occurs on the higher-order magnetic correlation. This is fundamentally different from the traditional thermodynamic phase transitions, in which the diverging spatial correlation length $\xi$ in the microscopic correlator $\langle (\hat{\sigma}_i^{x}-\langle\hat{\sigma}_i^{x}\rangle)(\hat{\sigma}_{i+\xi}^{x}-\langle\hat{\sigma}_{i+\xi}^{x}\rangle)\rangle$ leads to the divergence of the magnetic susceptibility~\cite{kardar2007statistical}. However, in the LMG model, the qubits are all coupled with each other with homogeneous strength and the qubits are indistinguishable. Thus, we cannot define a simple correlation length $\xi$ for the LMG model. Alternatively, we define a higher-order correlation function
\begin{equation}
C_{xxyy}=\frac{1}{2}\langle\hat{S}_x^2\hat{S}_y^2+\hat{S}_y^2\hat{S}_x^2\rangle_0 -\langle\hat{S}_x^2\rangle_0 \langle\hat{S}_y^2\rangle_0 \propto |B_x|^{-\tilde{\nu}},   
\end{equation}
to characterize the macroscopic correlation between the magnetic fluctuations in $x$- and $y$- axis.

\begin{figure}
\centering
\includegraphics[width=8.5cm]{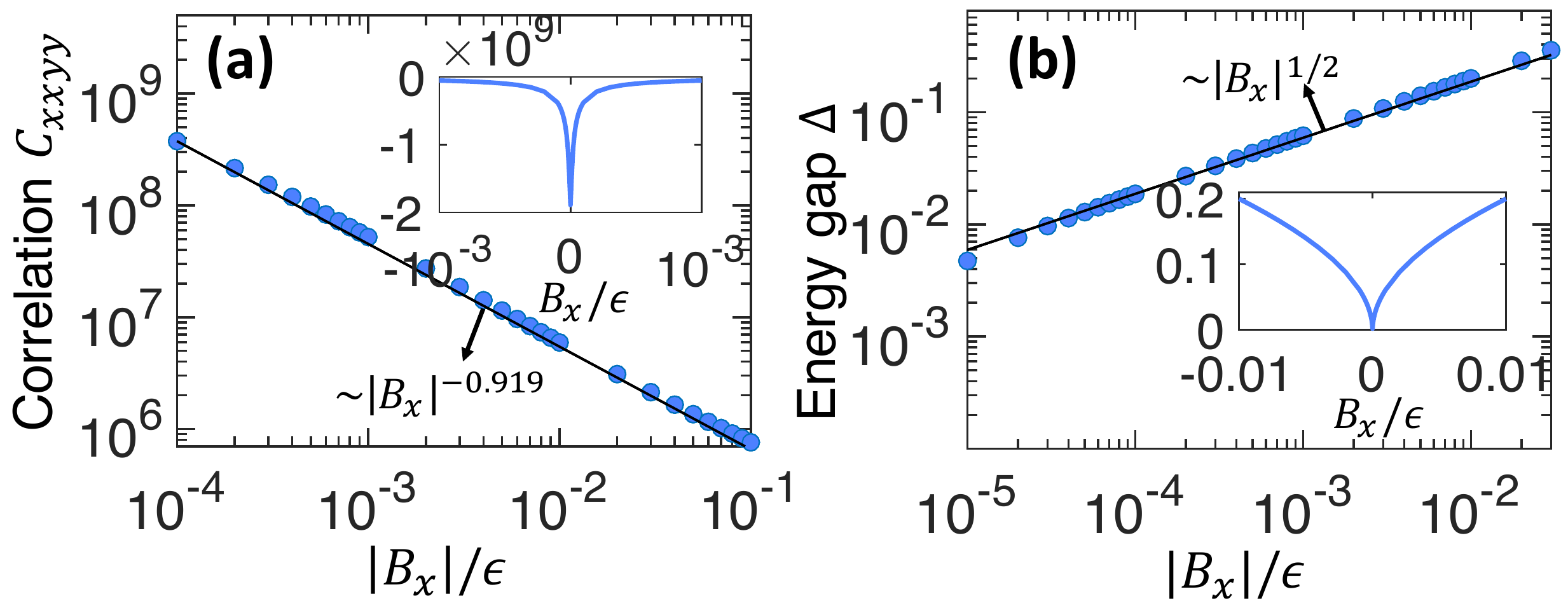}\caption{\label{fig:5} Singular behavior of the higher-order correlation and energy gap. \textbf{a} The higher-order correlation $C_{xxyy}$ diverges at the phase transition point $J_x=J_y=0.7\epsilon$. \textbf{b} The energy gap vanishes at the phase transition point. The subgraphs show the same curves in the linear coordinates. Both curves are symmetric on the two sides of the phase transition. The solid black lines are the algebraic fittings. The qubit number is set as $N=1000$.}
\end{figure}

The diverging $C_{xxyy}$ in the subgraph of Fig.~\ref{fig:5} (a) shows the strong negative correlation between $\hat{S}_x^2$ and $\hat{S}_y^2$ at the phase transition point. The negative correlation reveals the fact that the order parameter $\zeta_y$ decreases as the other one $\zeta_x$ increases. The corresponding singular exponent is $\tilde{\nu}\approx 0.919$ as shown by the black fitting curve. This exponent is universal for the LMG model, as it is independent on the qubit number $N$ as well as the position on the first-order QPT boundary in Fig.~\ref{fig:2}. We note that $\tilde{\nu}$ is similar to the traditional correlation length critical exponent~\cite{kardar2007statistical,dziarmaga2010dynamics}. We also find that the lower-order correlation $(1/2)\langle\hat{S}_x\hat{S}_y+\hat{S}_y\hat{S}_x\rangle_0-\langle\hat{S}_x\rangle_0\langle\hat{S}_y\rangle_0$ shows no singularity~\cite{supplementary}.

Another typical character of QPTs is that the energy gap $\Delta$ vanishes at the phase transition point as shown in Fig.~\ref{fig:5}b. The corresponding exponent is given by $\Delta\sim |B_x|^{1/2}$, which is same as the second-order QPT in LMG model~\cite{Defenu2018dynamical,Xue2018universal,Ribeiro2008exact}. Previous study on the size scaling for the LMG model shows that energy gap $\Delta$ also vanishes with the increasing qubit number $\Delta\sim 1/N$ at the phase transition point~\cite{Botet1983large,Dusuel2005continuous}.

\textbf{Conclusion}\\ 
Our work demonstrates the single-photon pulse induced giant response near a first-order QPT point. Our theoretical proposal can be directly implemented in current QPT simulators~\cite{zhang2017observation,bernien2017probing,Harris2018Phase}. We note that for microscopic systems, zero temperature generally implies preparing a system in a pure quantum (ground) state. We note however that in principle strongly interacting engineered qubits can show QPT behavior at finite temperature environments if the phase transition completes before the decoherence of the system occurs.

Our work paves a way for single-photon detection using quantum phase transition. This defect-controlled-QPT system is based on the fact that the first-order QPT in interaction qubit systems can be induced by a weak in-plane magnetic field. Future work will explore practical implementations on a circuit QED, cold atom and ion trap systems.

{\bf ACKNOWLEDGEMENTS}\\
This work is supported by the DARPA DETECT ARO award (W911NF-18-1-0074). 

{\bf AUTHOR CONTRIBUTIONS}\\
All authors contributed equally to this work.

{\bf COMPETING INTERESTS}\\
The authors declare no competing interests.

{\bf ADDITIONAL INFORMATION}\\
Supplementary information is available for this paper at~\cite{supplementary}.

{\bf DATA AVAILABILITY} \\
The data that support this study are available at https://github.com/yanglp091/QuantumPhaseTransitionDynamics

{\bf CODE AVAILABILITY}\\
The code that supports this study is available at https://github.com/yanglp091/QuantumPhaseTransitionDynamics

\bibliography{main}
\end{document}


\preprint{APS/123-QED}

\title{Dynamics of quantum phase transition and diverging higher-order correlation}

\author{Liping Yang}
\author{Zubin Jacob}
\email{zjacob@purdue.edu}
\affiliation{Birck Nanotechnology Center and Purdue Quantum Center, School of Electrical
and Computer Engineering, Purdue University, West Lafayette, IN 47906,
U.S.A.}

\maketitle

In this supplementary material, we first give the detailed absorption process in the defect center and the dynamics of this single photon triggered first-order quantum phase transition. Then, we explain the relation between the giant magnetic susceptibility and the diverging higher-order correlation in the large-number qubit system.

\section{Lambda-Transduction in NV Center \label{sec:lambda_transition}}

\begin{figure}
\centering
\includegraphics[width=8cm]{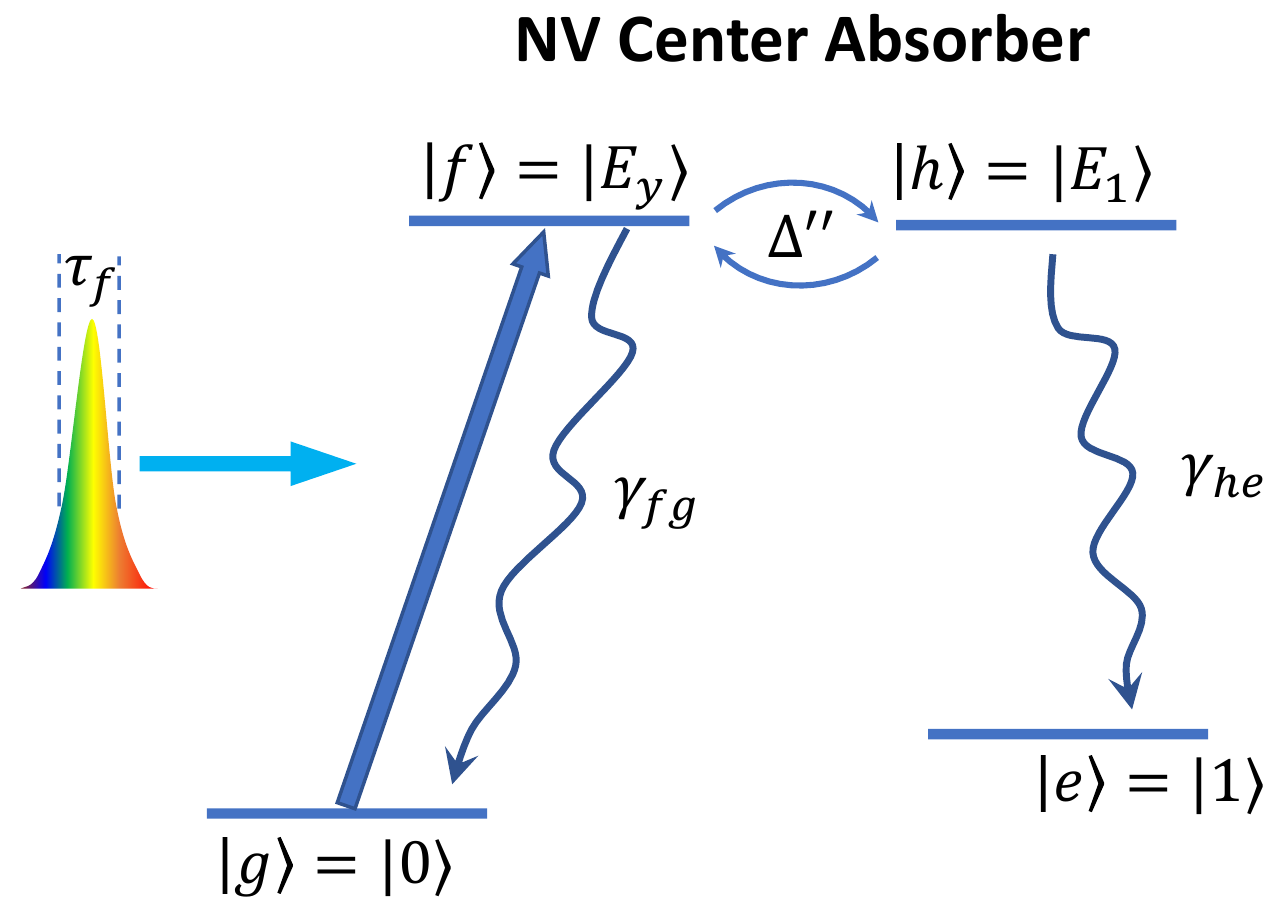}
\caption{\label{fig:S0}  Single-photon transduction (absorption) process. The incident single-photon pulse excited the NV center to the excited state $|f\rangle=|E_y\rangle$. Via the spin non-conservation coupling $\Delta^{\prime\prime}$ between states $|E_y\rangle$ and $|h\rangle =|E_1\rangle$, this excitation can be transferred to the ground spin state $|e\rangle=|1\rangle$ after the spontaneous decay. To realize an efficient transduction, the bandwidth matching between pulse length $\tau_f$, the spontaneous decay rates $\gamma_{fg}$ and $\gamma_{he}$, and the coupling strength $\Delta^{\prime\prime}$ must be carefully considered (see follow).}
\end{figure}

Usually, the optical transition in nitrogen-vacancy (NV) center does not change the state of the spin degree of freedom. Thus, a single-photon pulse (SPP) cannot induced a $\Lambda$-type transition to realize the flip of the two ground spin states $\left|0\right\rangle \rightarrow\left|1\right\rangle $ in an NV center. To sovle this problem, we need to construct a spin non-conservation transition in NV. One approach is combining a linearly polarized laser
with an additional circularly polarized laser, which has been demonstrated to realized all-optical control of the NV ground-state spin~\cite{Chu2015all}.
Here, we use another method by utilizing the energy crossing in the excited states of the NV as shown in Fig.~\ref{fig:S0}. We add strain to the NV at the energy crossing point of the two excited states
$\left|E_{y}\right\rangle =\left|a_{1}e_{y}-e_{y}a_{1}\right\rangle \otimes\left|0\right\rangle $
and $\left|E_{1}\right\rangle =\left|E_{-}\right\rangle \otimes\left|-1\right\rangle -\left|E_{+}\right\rangle \otimes\left|+1\right\rangle $
(with $E_{\pm}=a_{1}e_{\pm}-e_{\pm}a_{1}$ and $e_{\pm}=\mp(e_{x}\pm ie_{y})$)~\cite{maze2011properties}.
Here, $\left\{ \left|a_{1}\right\rangle ,\left|e_{x}\right\rangle ,\left|e_{y}\right\rangle \right\} $
and $\left\{ \left|-1\right\rangle ,\left|0\right\rangle ,\left|+1\right\rangle \right\} $
are the orbital basis of the excited states and the triplet spin states
(the two-hole representation). The coupling
$\Delta^{\prime\prime}$ between $\left|E_{y}\right\rangle $ and
$\left|E_{1}\right\rangle $ realizes the spin non-conservation transition.
Now, the effective Hamiltonian for the NV reads
\begin{equation}
\hat{H}_{{\rm NV}}=\Delta_{{\rm gs}}\left|1\right\rangle \left\langle 1\right|+\omega_{y}\left|E_{y}\right\rangle \left\langle E_{y}\right|+\omega_{1}\left|E_{1}\right\rangle \left\langle E_{1}\right|+\Delta^{\prime\prime}(\left|E_{y}\right\rangle \left\langle E_{1}\right|+\left|E_{1}\right\rangle \left\langle E_{y}\right|),\label{eq:HNV}
\end{equation}
where the energe of the ground spin state $\left|0\right\rangle $
is set as zero, $\Delta_{{\rm gs}}\approx2.87$~GHz the zero-field
splitting between ground spin states, and $\omega_{y}=\omega_{1}$
(the strain has been taken into account) are the energy difference
between the two excited states and the ground state $\left|0\right\rangle $. 

The interaction between the NV and the incident SPP is given by
\begin{equation}
\hat{H}_{{\rm pump}}=i\hbar\int_{0}^{\infty}d\omega[g(\omega)e^{i\vec{k}\cdot\vec{r}_{{\rm NV}}}\hat{a}(\omega)|E_{y}\rangle\left\langle 0\right|-{\rm h.c.}],\label{eq:Hint}
\end{equation}
where $\hat{a}(\omega)$ is the bosonic operator of the pulse mode
with frequency $\omega=c|\vec{k}|$, $\vec{r}_{{\rm NV}}$ is the
position of the NV center, and the rotating-wave approximation has
been taken. The amplitude of the NV-SPP interaction spectrum is given
by 
\begin{equation}
g(\omega)=\sqrt{\frac{\omega}{4\pi\varepsilon_{0}\hbar c\mathcal{A}}}(\vec{\epsilon}\cdot\vec{d}_{0y}),
\end{equation}
where $\mathcal{A}$ is the effective transverse cross section of
the pulse~\cite{yang2018concept}, $\vec{d}_{0y}$ the electric dipole
vector of the $\left|0\right\rangle \rightarrow\left|E_{y}\right\rangle $
transition, and the unit vector $\vec{\epsilon}$ denotes the polarization
of the pulse. The wave-packet amplitude of a Gaussian SPP is given
by 
\begin{equation}
\xi(t)=\left(\frac{1}{2\pi\tau_{f}^{2}}\right)^{1/4}\exp\left[-\frac{t^{2}}{4\tau_{f}^{2}}-i\omega_{0}t\right],  \label{eq:pulsewave}  
\end{equation}
with center frequency $\omega_{0}$ and pulse length $\tau_{f}$~\cite{yang2018concept}.
The incident SPP pulse is resonant with $\left|0\right\rangle \rightarrow\left|E_{y}\right\rangle $
transition, i.e., $\omega_{0}=\omega_{y}=\omega_{1}$.

We note that the excited state $\left|E_{1}\right\rangle $ is a superposition
of states with spin $\left|-1\right\rangle $ and $\left|+1\right\rangle $.
After the spontaneous decay, a quantum entanglement state $\left|\psi\right\rangle =(|\sigma_{-}\rangle|-1\rangle-|\sigma_{+}\rangle|+1\rangle)$
between a outgoing circularly polarized single photon and the ground spin states of the NV is  obtained~\cite{togan2010quantum}. The amplifier performs a projection measurement on the spin state of the NV. Each time, only one of the spin states can be detected. Actually, both of the ground spin states $\left|\pm1\right\rangle $ can provide an effective magnetic field (with inverse direction) for amplifier spins to trigger the quantum phase transition. If the coherence of the NV center has been preserved during the amplification process, the whole system will finally go to a NV-amplifier entangled state, i.e., $(|+1\rangle\otimes |-M_x\rangle + |-1\rangle\otimes |M_x\rangle)/\sqrt{2}$ ($|\pm M_x\rangle$ are the excites states of the amplifier with positive and negative spontaneous magnetization in $x$-axis, respectively). Here, without loss of generality, we only take the case that the NV decays to the ground spin state $\left|1\right\rangle $ as an example. For simplicity,
we use the following denotation hereafter
\begin{align}
\left|0\right\rangle  & =\left|g\right\rangle \\
\left|1\right\rangle  & =\left|e\right\rangle \\
\left|E_{y}\right\rangle  & =\left|f\right\rangle \\
\left|E_{1}\right\rangle  & =\left|h\right\rangle 
\end{align}

\begin{figure}
\centering
\includegraphics[width=8cm]{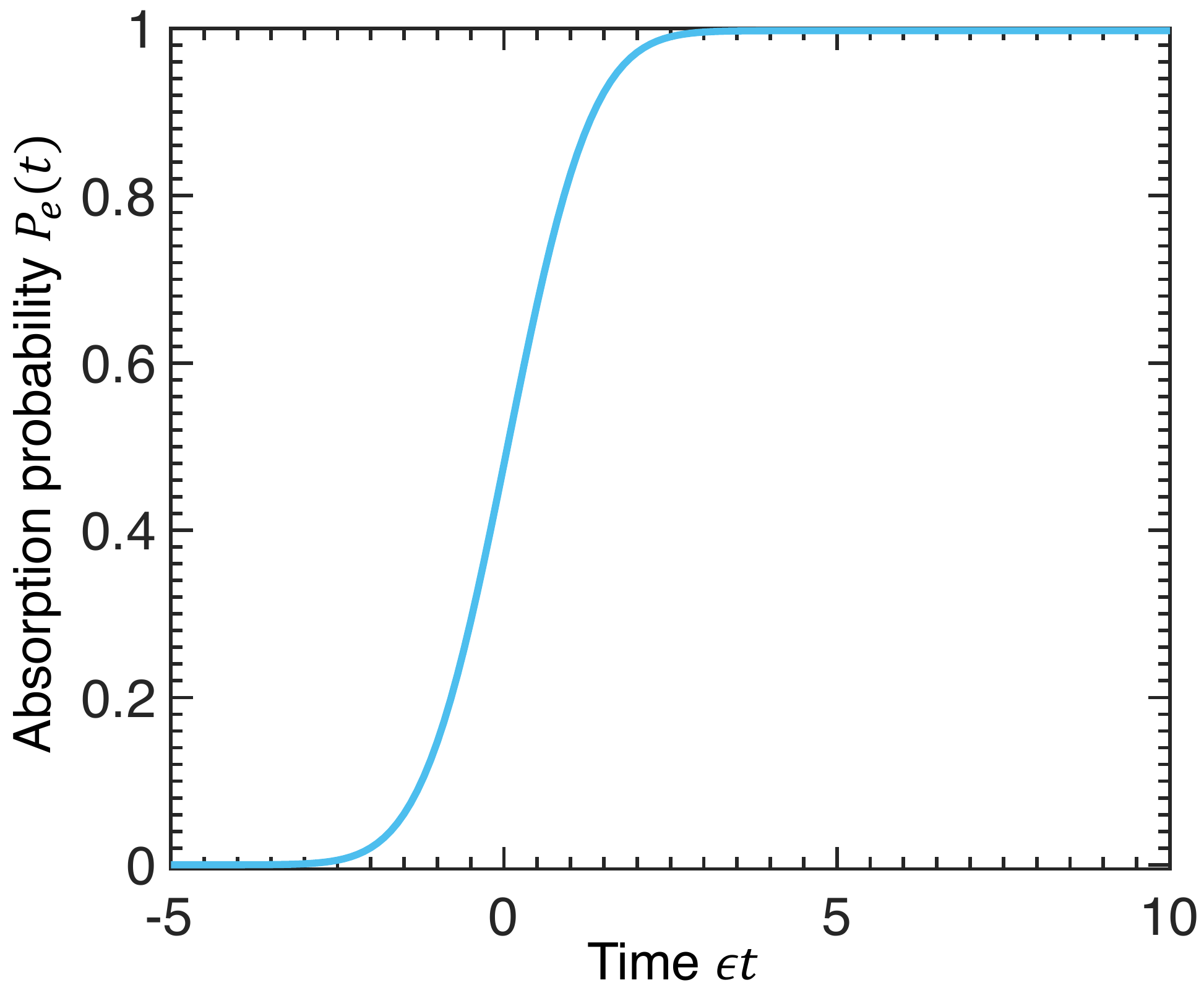}
\caption{\label{fig:S1} \textbf{Time-dependent absorption probability.} The length of the Gaussian pulse is set as $\epsilon\tau_f=1$. The coupling between the two excited state $|f\rangle$ and $|h\rangle$ is $\Lambda^{\prime\prime}=5\epsilon$ and the spontaneous decay rates of these two excited states are $\gamma_{fg}=\gamma_{he}=\Gamma=10\epsilon$. }
\end{figure}

\section{Dynamics of Single-Photon interacting with a large quantum system \label{sec:dynamics}}

The full dynamics of the whole system under the pumping of the center
spin by a single-photon pulse is given by a time-dependent master equation~\cite{yang2019single,Baragiola2012n-photon}
\begin{equation}
\frac{d}{dt}\rho_{{\rm tot}}(t)=-i[\hat{H},\rho_{{\rm tot}}(t)]+\mathcal{L}_{{\rm P}}(t)\rho_{{\rm tot}}(t)+\mathcal{L}_{{\rm SD}}\rho_{{\rm tot}}(t).\label{eq:MEQ2-1}
\end{equation}
Here, $\hat{H}=\hat{H}_{{\rm NV}}+\hat{H}_{{\rm Am}}+\hat{H}_{{\rm int}}$ is the Hamiltonian of the whole system. The Hamiltonian of the NV is given in equation (\ref{eq:HNV}). The the amplifier is described by the Lipkin-Meshkov-Glick (LMG) model~\cite{lipkin1965validity,meshkov1965validity,glick1965validity},
\begin{equation}
\hat{H}_{{\rm Am}}=\frac{1}{2}\epsilon\sum_{j=1}^{N}\hat{\sigma}_{j}^{z}-\frac{1}{N}\sum_{i<j}(J_{x}\hat{\sigma}_{i}^{x}\hat{\sigma}_{j}^{x}+J_{y}\hat{\sigma}_{i}^{x}\hat{\sigma}_{j}^{x})
\end{equation}
with energy splitting $\epsilon$ in $z$-direction and the homogeneous long-range couplings $J_x$ and $J_y$ in $xy$-plane. The interaction between the NV absorber and the amplifier qubits is described by
\begin{equation}
\hat{H}_{\rm int}=B_x \left|e\right\rangle \left\langle e\right|\sum_{j}\hat{\sigma}_{j}^{x}.
\end{equation}

The initial density matrix $\rho_{{\rm tot}}(t_0)=I_{p}\otimes\rho_{{\rm NV}}(t_0)\otimes\rho_{{\rm Am}}(t_0)$ of the whole system is composed of three parts: (1)
 $I_{p}$ is the $2\times 2$ identity matrix for a $n$-photon Fock-state pulse; (2) $\rho_{{\rm NV}}(t_0)=\left|g\right\rangle \left\langle g\right|$ for the ground-state NV center; (3) $\rho_{{\rm Am}}(t_0)$ the ground-state of $H_{{\rm Am}}$ with specifically engineered bias couplings $J_x$ and $J_y$.

The pumping from a quantum pulse is given by 
\begin{equation}
\mathcal{L}_{{\rm P}}\rho_{{\rm tot}}=\sqrt{\gamma_{fg}}\eta\left\{ \xi(t-t_{0})e^{i\phi}[\hat{\tau}_{+}\rho_{{\rm tot}},\hat{\sigma}_{fg}]+\xi^{*}(t-t_{0})e^{-i\phi}[\hat{\sigma}_{gf},\rho_{{\rm tot}}\hat{\tau}_{-}]\right\} ,
\end{equation}
with the spontaneous decay rate $\gamma_{fg}$ from the excited state $|f\rangle$ back to the ground state $|g\rangle$ and ladder operators $\hat{\sigma}_{fg}=\left|f\right\rangle \left\langle g\right|$ and $\hat{\sigma}_{gf}=\left|g\right\rangle \left\langle f\right|$. Here, $\eta_{j}$ characterizes the scattering efficiency of the NV, $\phi=\vec{k}_{0}\cdot\vec{r}_{{\rm NV}}$ ($\omega_{0}=c|\vec{k}_{0}|$), $t_0$ is the propagating time for pulse to arrive at the NV center, and the time-dependent wave-packet amplitude $\xi(t)$ of the Fock-state pulse is given in equation (\ref{eq:pulsewave}). The raising
operator $(\hat{\tau}_{-})^{\dagger}=\hat{\tau}_{+}=\left[\begin{array}{cc}
0 & 1\\
1 & 0
\end{array}\right]$couples the different photon-number subspace for Fock-state pulse. 

\begin{figure}
\includegraphics[width=8cm]{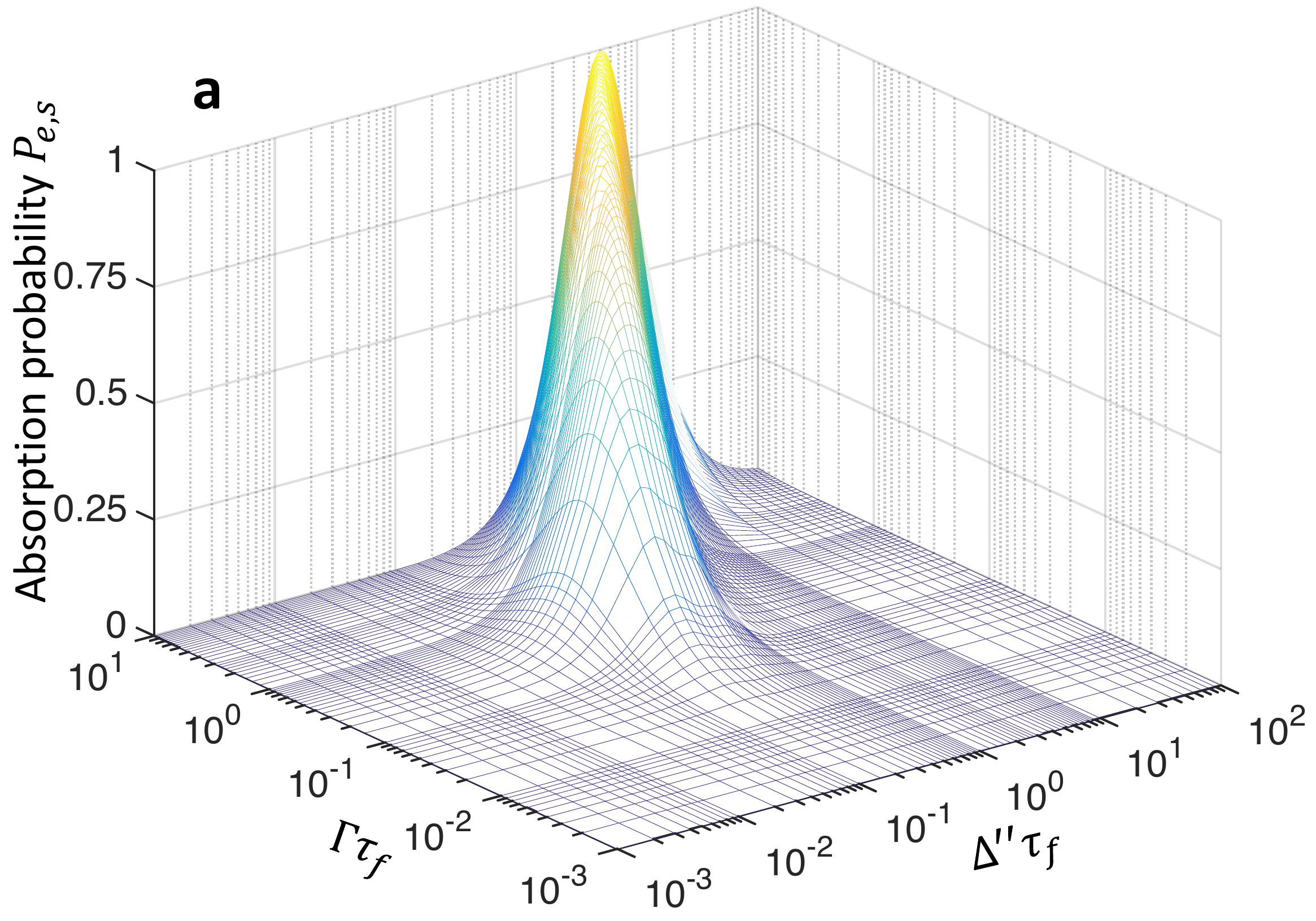}
\includegraphics[width=8cm]{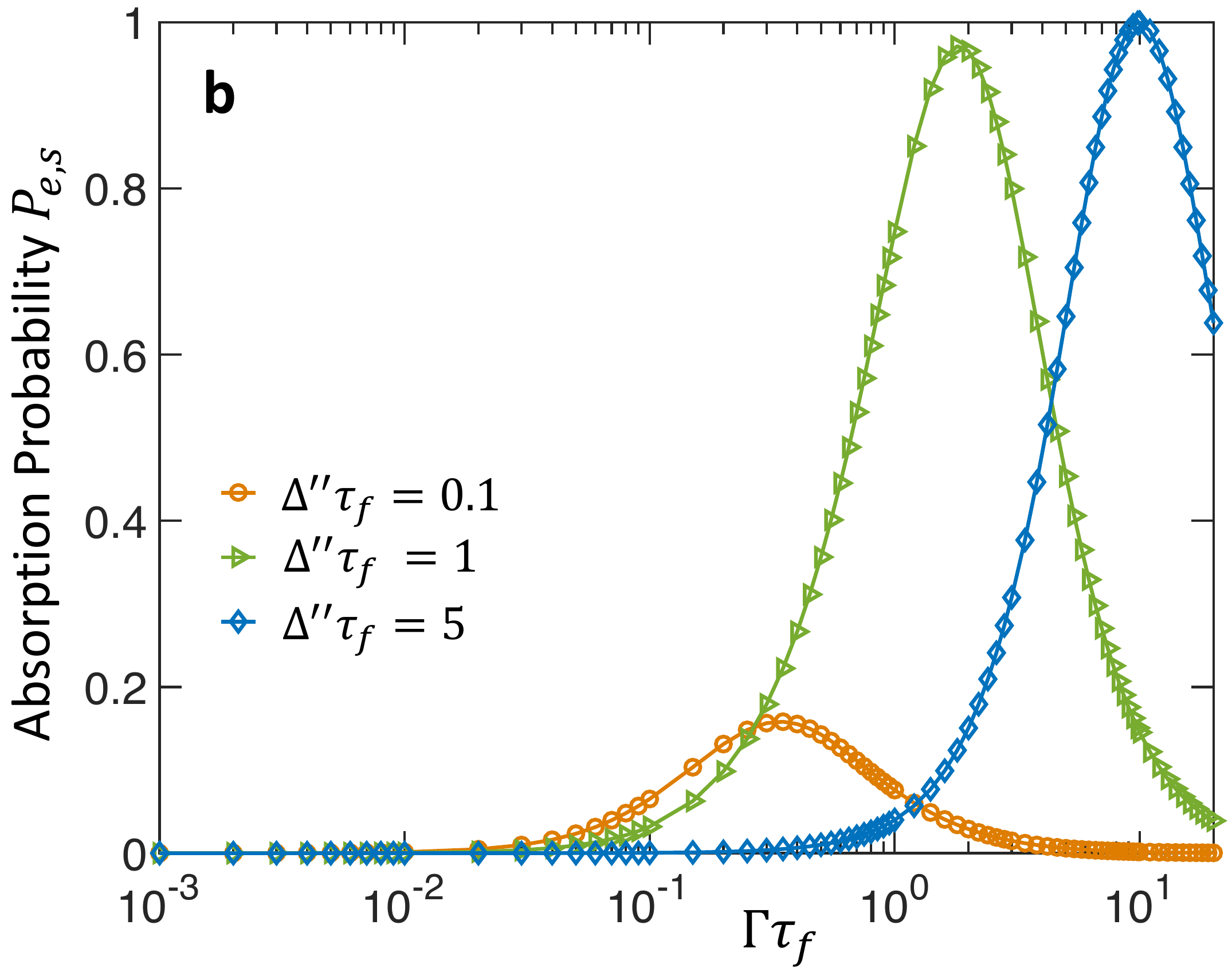}
\caption{\label{fig:S2} \textbf{Optimization of the transduction probability. a} The steady-state (the max) value $P_{e,s}$ of the population of the state $|e\rangle$ as a function of $\Delta^{\prime\prime}$ and $\Gamma$ is shown. Here, $\Delta^{\prime\prime}$ is the coupling between the two excited states $|f\rangle$ and $|h\rangle$. The two spontaneous decay rates are set to be the same $\gamma_{fg}=\gamma_{he}=\Gamma$. The pulse length is set as $\tau_f\epsilon =1$. \textbf{b} The steady-state probabilities $P_{e,s}$ for specific couplings $\Delta^{\prime\prime}$ are given. }
\end{figure}

As usually the coherence time of the amplifier is much longer than amplification time, we only consider the decay of the center spin from the electronic excited state
\begin{align}
\mathcal{L}_{{\rm SD}}\rho_{{\rm tot}} & =\gamma_{fg}[\hat{\sigma}_{gf}\rho_{{\rm tot}}\hat{\sigma}_{fg}-\frac{1}{2}\hat{\sigma}_{fg}\hat{\sigma}_{gf}\rho_{{\rm tot}}-\frac{1}{2}\rho_{{\rm tot}}\hat{\sigma}_{fg}\hat{\sigma}_{gf}]\\ \nonumber
&+\gamma_{he}[\hat{\sigma}_{eh}\rho_{{\rm tot}}\hat{\sigma}_{he}-\frac{1}{2}\hat{\sigma}_{he}\hat{\sigma}_{eh}\rho_{{\rm tot}}-\frac{1}{2}\rho_{{\rm tot}}\hat{\sigma}_{he}\hat{\sigma}_{eh}].
\end{align}

We note that  due to the dispersive coupling between the NV center and the amplifier, the dynamics of the absorber and the amplifier are almost "decoupled". After trace off the NV degree of freedom, the dynamics of the amplifier
is given by
\begin{equation}
\frac{d}{dt}\rho_{{\rm Am}}(t)=-i[\hat{H}_{{\rm Am}}^{\prime},\rho],
\end{equation}
where
\begin{equation}
\hat{H}_{{\rm Am}}^{\prime}=\hat{H}_{{\rm Am}}+P_{e}(t)B_{x}\sum_{j}\hat{\sigma}_{j}^{x},
\end{equation}
and $P_{e}(t)$ is the population of the NV in the state $\left|e\right\rangle $.
Here, we see that after transduction, the amplifier qubits experience an effective magnetic field $P_e(t)\times B_x$ from the absorber. The dynamics of the absorber and the amplifier can be evaluated separately. 

In Fig.~\ref{fig:S1}, we give the time-dependent population (the net absorption probability) $P_e(t)$ of the state $|e\rangle$. The absorption probability $P_e$ increases after the pulse arrives and finally reaches a steady-state value $P_{e,s}$. Here, the dissipation of the state back to $|g\rangle$ has been neglected due to the long life time of the metastate $|e\rangle$. 

To realize an efficient single-photon transduction, we need to optimize the pulse length $\tau_f$, the coupling strength $\Delta^{\prime\prime}$ between the two excited states, and the two spontaneous decay rates $\gamma_{fg}$ and $\gamma_{he}$. In experiment, we can use filters to tailor the pulse spectrum and change the pulse length~\cite{weiner1995femtosecond}. The typical value of the coupling strength $\Delta^{\prime\prime}$ is around $~1$~GHz. Usually, it is hard to tune this coupling strength. However, we can engineer the density of state of the electromagnetic fields to tune the decay rates $\gamma_{fg}$ and $\gamma_{he}$~\cite{Purcell1946resonance,Kleppner1981inhibited,Yablonovitch1987inhibited} to enhance the transduction efficiency. In Fig.~\ref{fig:S2}a, we show the optimization conditions for larger transduction probability. It has been shown that nearly unit transduction probability $P_{e,s}$ can be realized when $\gamma_{fg}=\gamma_{he}=\Gamma\gg 1/\tau_f$ for three-level atom system~\cite{young2018fundamental}. In Fig.~\ref{fig:S2}b, we show that for four-level systems, unit probability $P_{e,s}$ can also be obtained  when $\Gamma=2\Delta^{\prime\prime}\gg 1/\tau_f$.

\begin{figure}
\centering
\includegraphics[width=9cm]{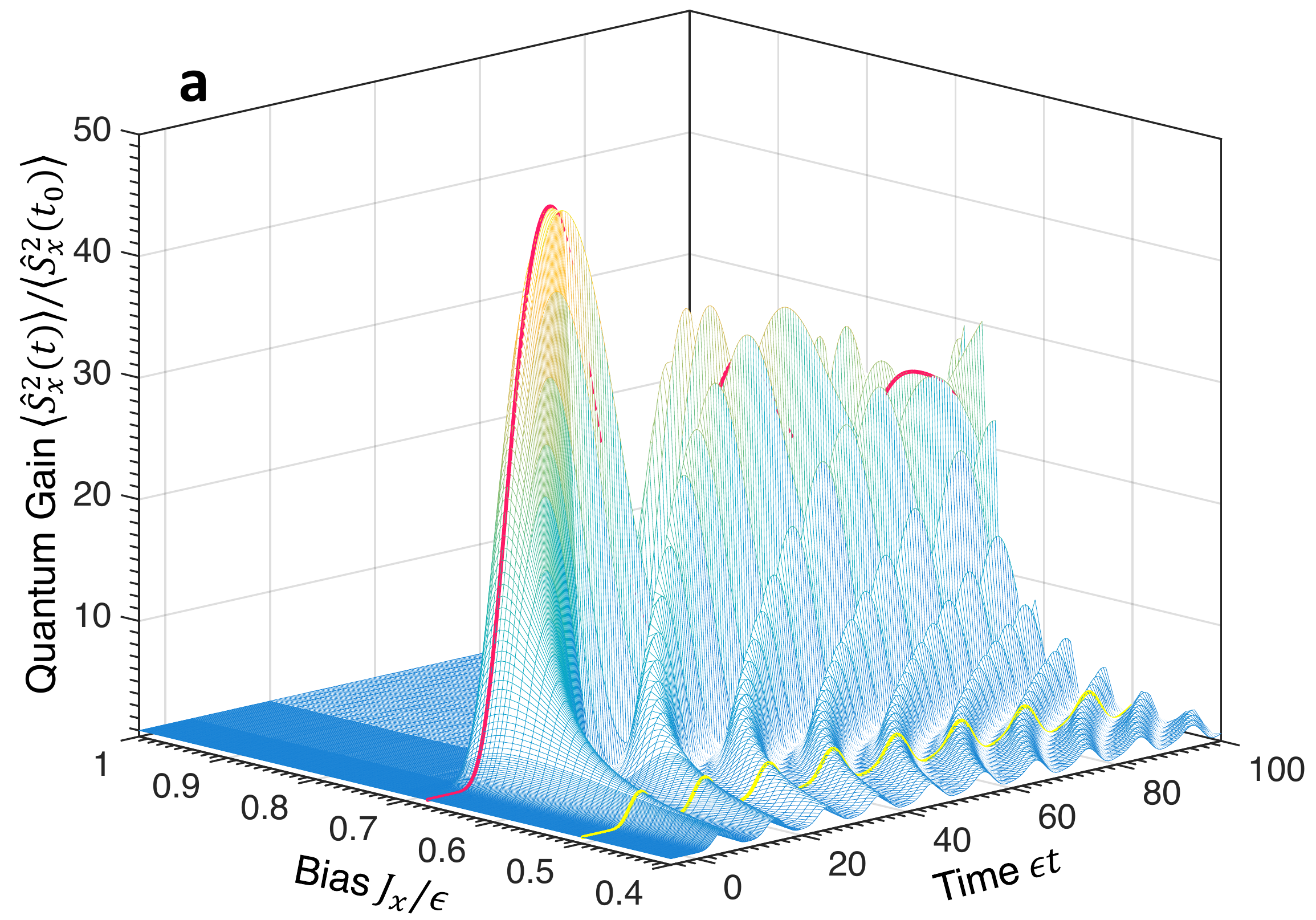}
\includegraphics[width=8cm]{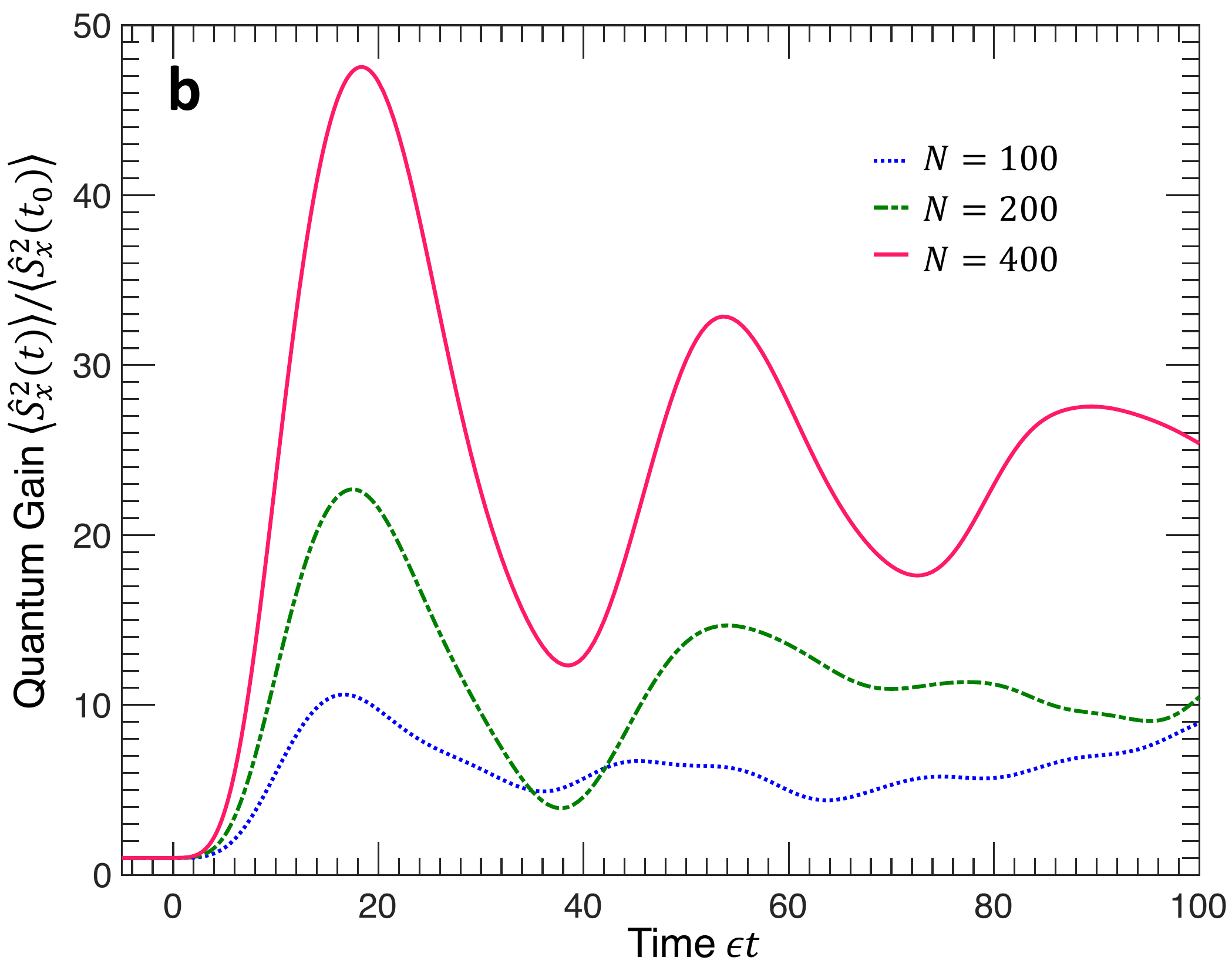}
\caption{\label{fig:S3} \textbf{Single-photon pulse induced first-order quantum phase transition. a} The quantum gain as a function of time and the bias qubit-qubit coupling $J_x$ is shown. After absorption of a single-photon pulse, the absorber is flipped to the state $|e\rangle$, on which the absorber exerts a weak magnetic field $B_x \times P_e(t)$ ($ P_e(t)$ the population of the state $|e\rangle$) on the amplifier with $N = 400$ qubits. Only if the the amplifier is optically biased around the critical point $J_{x,c} \equiv J_y$, the time varying field can trigger a first-order quantum phase transition to obtain a large quantum gain. Here, the qubit-qubit coupling in $y$-direction is fixed at $J_y = 0.7\epsilon$ and the absorber-amplifier coupling $B_x = 0.01\epsilon$. \textbf{b} The time-dependent quantum gain $G(t)$ for different qubit-number is shown. Here, the qubit-qubit couplings are $J_x=0.675\epsilon$ and $J_y=0.7\epsilon$. }
\end{figure}

In Fig.~\ref{fig:S3}a, we show that only if the qubit-qubit coupling $J_x$ is biased close to the phase transition point $J_{x,c}=J_y$, a large quantum gain can be obtained. The red curve and the yellow curve correspond to $J_x=0.675\epsilon$ and $J_x=0.5\epsilon$, respectively. In Fig.~\ref{fig:S3}b, we show that time $T_{\rm Am}$ to reach the maximum of the quantum gain is almost independent on the qubit number $N$. However, the quantum gain increases with $N$ linearly, which is consistent with our previous result~\cite{yang2019QCD}

\section{Higher-Order Correlation\label{sec:higherCorrelation}}
In thermodynamic phase transitions, the divergences of the magnetic susceptibility and spacial correlation length are directly related. The Gibbs partition function in a magnetic field $h$ is given by~\cite{kardar2007statistical}
\begin{equation}
Z={\rm Tr}\exp(-\beta \hat{H}_{0}+\beta h_{\alpha}\hat{M}_{\alpha}),
\end{equation}
where $\hat{H}_{0}$ describes the internal energy of the magnet including qubit-qubit interactions and $-h_{\alpha}\hat{M}_{\alpha}$
is the work done against the magnetic field to produce a magnetization
$\langle\hat{M}_{\alpha}\rangle$ in the direction $\alpha = x,y,{\rm or}\ z$. The equilibrium magnetization is computed from
\begin{equation}
\langle \hat{M}_{\alpha}\rangle=\frac{\partial\ln Z}{\partial\beta h_{\alpha}}=\frac{1}{Z}{\rm Tr}[\hat{M}\exp(-\beta \hat{H}_{0}+\beta h_{\alpha}\hat{M}_{\alpha})],
\end{equation}
and the susceptibility is then related to the variance of the magnetization by
\begin{align}
\chi_{\alpha} & =\frac{\partial \langle\hat{M}_{\alpha}\rangle}{\partial h_{\alpha}}\\
& = \beta\left\{ \frac{1}{Z}{\rm Tr}[\hat{M}_{\alpha}^{2}\exp(-\beta \hat{H}_{0}+\beta h_{\alpha}\hat{M}_{\alpha})]-\frac{1}{Z^{2}}{\rm Tr}[\hat{M}_{\alpha}\exp(-\beta \hat{H}_{0}+\beta h_{\alpha}\hat{M}_{\alpha})]^{2}\right\} \\
 & =\frac{1}{k_{B}T}\left(\langle \hat{M}_{\alpha}^{2}\rangle-\langle \hat{M}_{\alpha}\rangle^{2}\right).
\end{align}

\begin{figure}
\centering
\includegraphics[width=8cm]{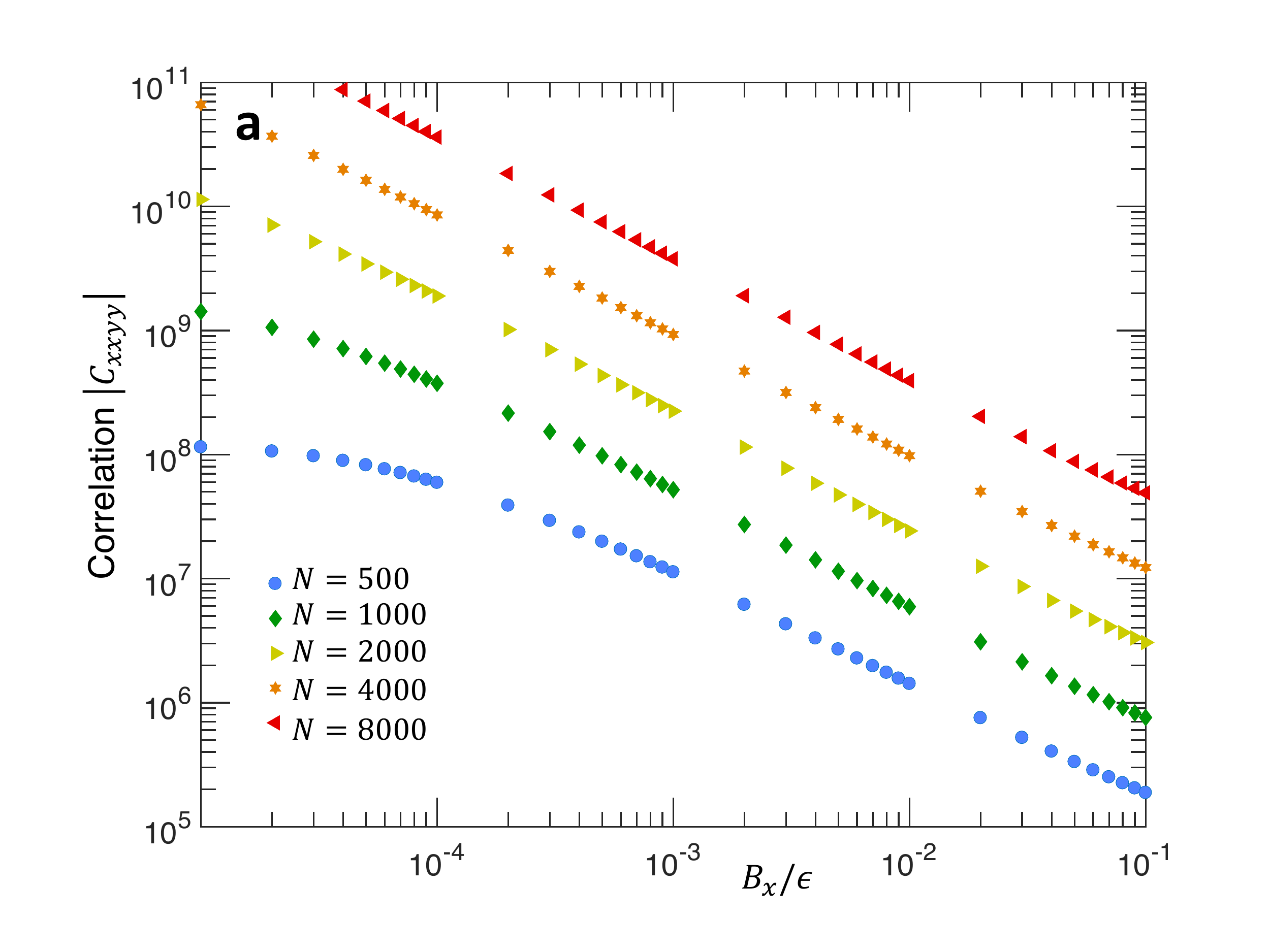}
\includegraphics[width=8cm]{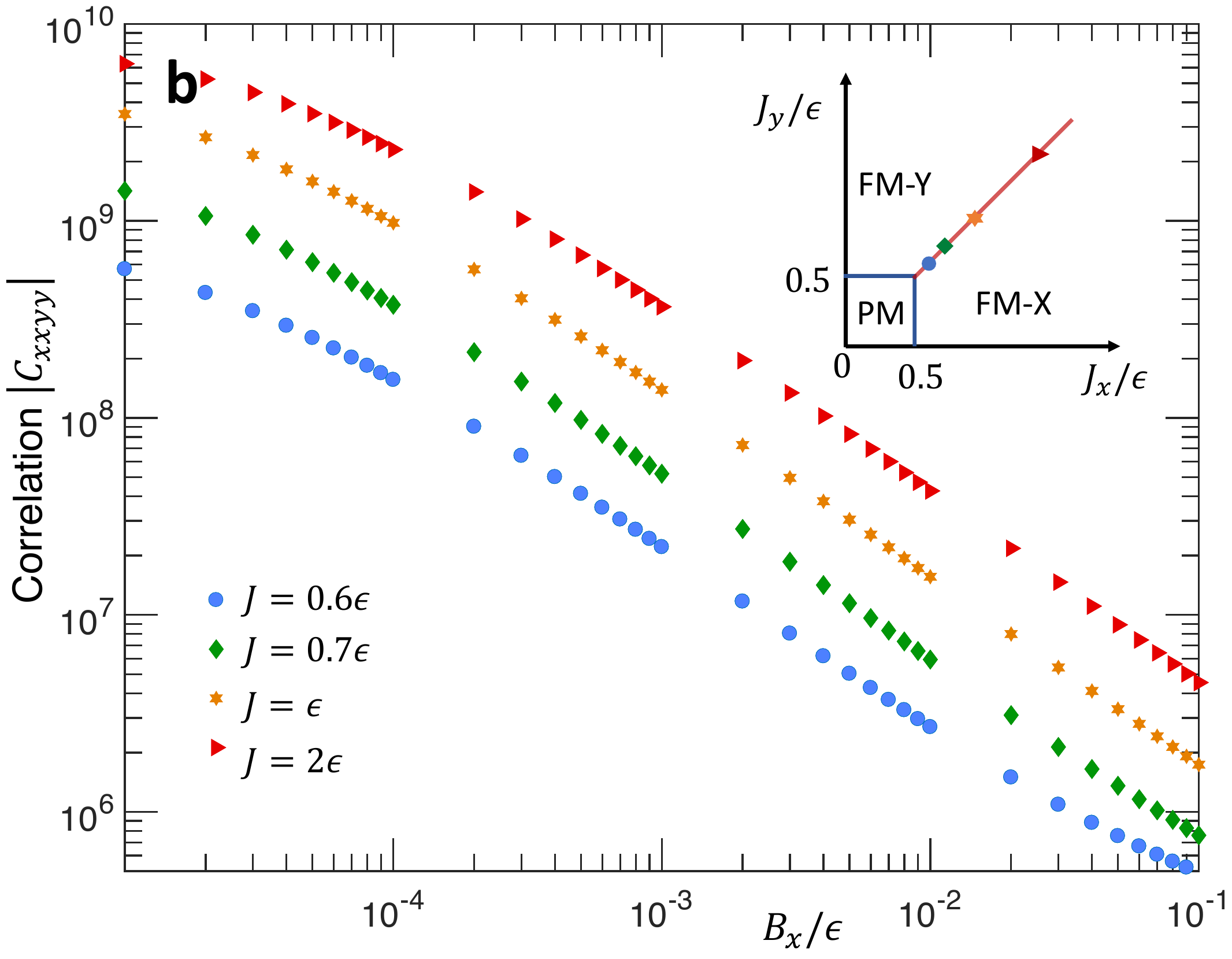}
\caption{\label{fig:S6} \textbf{Universal singular exponent. a} The correlation functions $C_{xxyy}$ for different qubit number $N$ are shown. Here, the qubit-qubit coupling is fixed at $J_x=J_y=0.7/\epsilon$. \textbf{b} The correlation functions $C_{xxyy}$ for different qubit-qubit coupling strength $J_x=J_y=J$ are shown. The subgraph shows the corresponding positions on the phase diagram for different $J$. Here, the qubit number is fixed at $N=1000$. The singular scaling $C_{xxyy}\propto|B_x|^{-\gamma}$ does not change with qubit number $N$ as well as  $J$.   }
\end{figure}

\begin{figure}
\centering
\includegraphics[width=10cm]{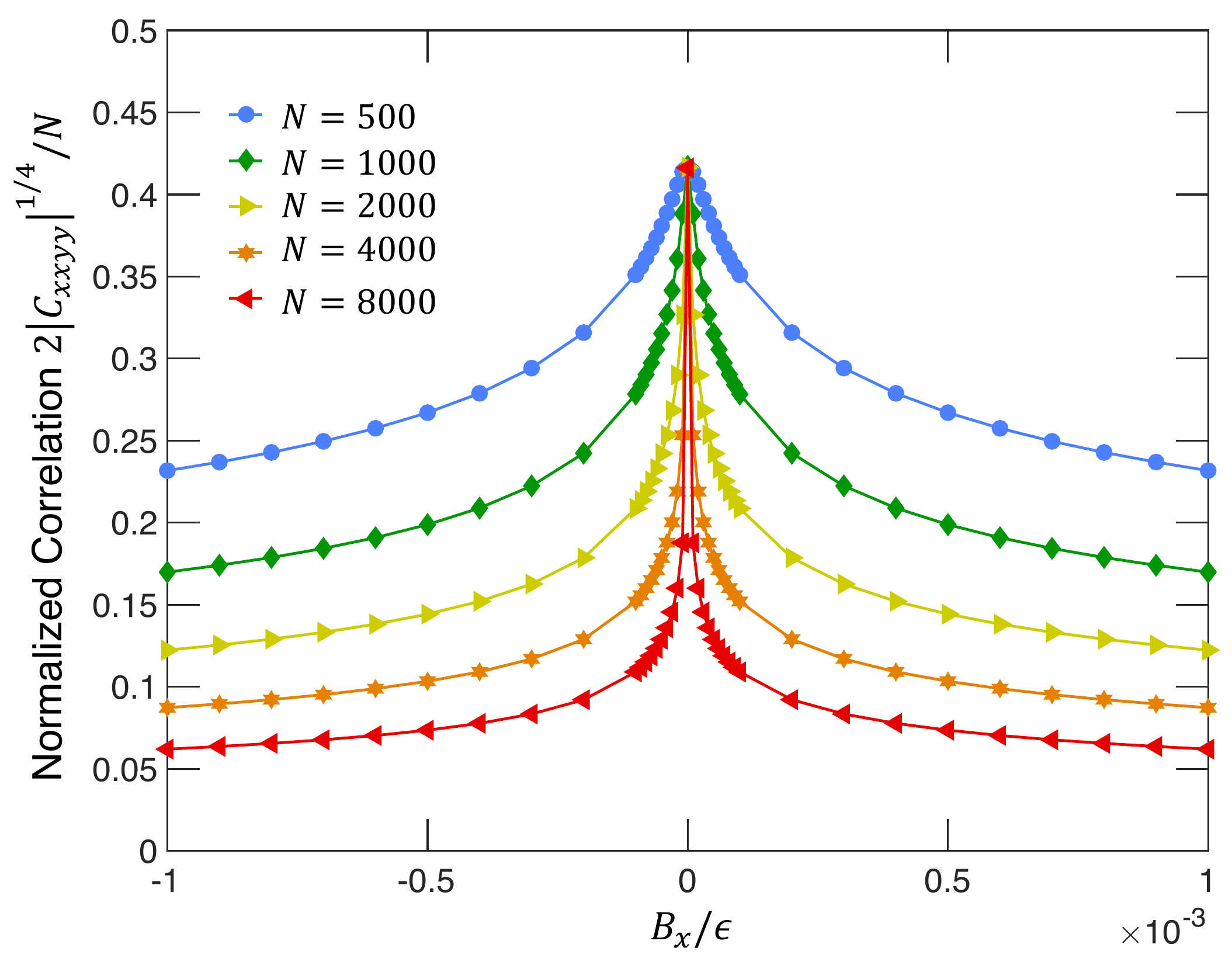}
\caption{\label{fig:S8} \textbf{Re-scaled correlation function.} Different curves denote different qubit number. Here, the qubit-qubit coupling is fixed at $J_x=J_y=0.7/\epsilon$.}
\end{figure}

The magnetization operator for discrete lattice systems is given by
\begin{equation}
\hat{M}_{\alpha}=\frac{1}{2}\sum_{j}\hat{\sigma}_{j}^{\alpha},
\end{equation}
for lattice system. Then, the relation between the susceptibility and the spatial correlation function is given by
\begin{equation}
\chi=\frac{1}{4k_{B}T}\sum_{ij}\left(\langle\hat{\sigma}_i^{\alpha}\hat{\sigma}_j^{\alpha}\rangle-\langle\hat{\sigma}_i^{\alpha}\rangle\langle\hat{\sigma}_j^{\alpha}\rangle\right)\equiv\frac{1}{k_{B}T}C_{\alpha\alpha}.\label{eq:chi-G}
\end{equation}
Utilizing the transnational symmetry of a homogeneous system, we can connect the bulk response function with the microscopic two point correlation functions,
\begin{align}
C_{\alpha\alpha} & =\frac{N}{4}\sum_j\langle[\hat{\sigma}_{j}^{\alpha}-\langle\hat{\sigma}_{j}^{\alpha}\rangle][\hat{\sigma}_{1}^{\alpha}-\langle\hat{\sigma}_{1}^{\alpha}\rangle]\rangle\equiv\frac{N}{4}\sum_{j}G_{1j}^{\alpha\alpha}.\label{eq:chi-G1}
\end{align}
In many cases, the correlation function decays as $G_{1j}^{\alpha\alpha}\propto\exp(-|j-1|/\xi)$ at separations $|j-1|>\xi$. Here,
$\xi$ called the correlation length is the only relevant length at the phase transition point.

However, at the first-order quantum phase transition (QPT) points, the singular behaviors occur on the correlation between the magnetic fluctuations in $x$ and $y$ directions. It can be easily verified that the lowest-order symmetrized macroscopic correlation function 
\begin{equation}
C_{xy}=\frac{1}{2}\langle\hat{S}_x\hat{S}_y+\hat{S}_y\hat{S}_x\rangle-\langle\hat{S}_x\rangle\langle\hat{S}_y\rangle,
\end{equation}
due to the symmetry of the spontaneous magnetization in $xy$-plane. Here, $\hat{S}_{\alpha}=\sum_i\hat{\sigma}_j^{\alpha}/2$ is the collective angular momentum operator. Thus, we need to consider the higher-order correlation 
\begin{equation}
C_{xxyy}=\frac{1}{2}\langle\hat{S}_x^2\hat{S}_y^2+\hat{S}_y^2\hat{S}_x^2\rangle-\langle\hat{S}_x^2\rangle\langle\hat{S}_y^2\rangle.
\end{equation}

The singular scaling of $C_{xxyy}$ has been shown in Fig.~4 (a) in the main text. Here, we show that this scaling is independent on the qubit number and the position on the phase transition boundary. In Fig.~\ref{fig:S6}a, we contrast the correlation $C_{xxyy}$ with different qubit number. In Fig.~\ref{fig:S6}b, we contrast the correlation $C_{xxyy}$ with different qubit-qubit coupling $J_x=J_y=J$. We see that the value of $C_{xxyy}$ changes, but the scaling exponent $\gamma$ of  $C_{xxyy}\propto |B_x|^{-\gamma}$ at the phase transition point remains the same.

As explained in the main text, we can not define a simple correlation length $\xi$ for the LMG model with indistinguishable  qubits. However, we may used the rescaled correlation function
\begin{equation}
\eta = \frac{2}{N}|C_{xxyy}|^{1/4},    
\end{equation}
to characterize the proportion of correlated qubits. As shown in Fig.~\ref{fig:S8}, the size of correlated qubit clusters decreases away from the phase transition point.

\bibliography{main}